\documentclass[twocolumn,nofootinbib,aps,preprintnumbers,amsmath,amssymb]{revtex4-1}
\usepackage{graphics}

\usepackage{graphicx}
\usepackage{amssymb}
\usepackage{amsbsy}
\usepackage{multirow}
\usepackage{mathrsfs}
\usepackage{amsmath}
\usepackage[T1]{fontenc}
\DeclareMathAlphabet{\mathscrbf}{OMS}{mdugm}{b}{n}
\usepackage{bm}
\usepackage{color}
\usepackage{epstopdf}
\usepackage[normalem]{ulem}
\usepackage{verbatim}
\usepackage[table]{xcolor}
\usepackage{float}
\restylefloat{table}
\usepackage{placeins}
\usepackage{booktabs}

\newcommand{\tauplum}{\mathcal{T}^{\rm P}}
\newcommand{\tausb}{\mathcal{T}^{\rm S}}

\newcommand{\beq}{\begin{equation}}
\newcommand{\eeq}{\end{equation}}

\newcommand{\Red}[1]{{\textcolor{red}{#1}}}

\newcommand{\Trm}{{\bf T}}

\newcommand{\Crm}{{\bf C}}

\newcommand{\At}{{\rm Ala}_{3}}
\newcommand{\Aq}{{\rm Ala}_{4}}
\newcommand{\Rof}{R_{\operatorname{1--4}}}
\newcommand{\of}{\operatorname{1--4}}

\newcommand{\JSDeone}{{\rm JSD}({\hat{\boldsymbol e}_1} \mid\mid {\hat{\boldsymbol e}_1}^{\operatorname{BMSM}})}
\newcommand{\JSDpi}{{\rm JSD}({\boldsymbol \pi} \mid\mid {\boldsymbol \pi}^{\operatorname{CG-sb}})}

\begin{document}
%
\title{Concurrent parametrization against static and kinetic information leads
  to more robust coarse-grained force fields}
 

\author{Joseph F.~Rudzinski}
\author{Tristan Bereau}
\affiliation{Max Planck Institute for Polymer Research, Mainz 55128, Germany}
\date{\today}

\begin{abstract} 
  The parametrization of coarse-grained (CG) simulation models for molecular
  systems often aims at reproducing static properties alone.  The reduced
  molecular friction of the CG representation usually results in faster,
  albeit inconsistent, dynamics. In this work, we rely on Markov state models
  to simultaneously characterize the static and kinetic properties of two CG
  peptide force fields---one top-down and one bottom-up.  Instead of a
  rigorous evolution of CG dynamics (e.g., using a generalized Langevin
  equation), we attempt to improve the description of kinetics by simply
  altering the existing CG models, which employ standard Langevin dynamics.
  By varying masses and relevant force-field parameters, we can improve the
  timescale separation of the slow kinetic processes, achieve a more
  consistent ratio of mean-first-passage times between metastable states, and
  refine the relative free-energies between these states.  Importantly, we
  show that the incorporation of kinetic information into a structure-based
  parametrization improves the description of the helix-coil transition
  sampled by a minimal CG model.  While structure-based models understabilize
  the helical state, kinetic constraints help identify CG models that improve
  the ratio of forward/backward timescales by effectively hindering the
  sampling of spurious conformational intermediate states.
\end{abstract}
\maketitle
\section{Introduction}
\label{intro}

The last few decades have witnessed the onset and development of computer
simulations of (macro)molecular systems, providing increasingly accurate and
reliable atomic-level detail into their structure and dynamics
\cite{van1990computer, Karplus:2002, lane2013milliseconds}.  For all but the
smallest of systems, sufficient conformational sampling remains a significant
bottleneck---notable examples include conformational transitions in proteins
\cite{lane2013milliseconds} and the insertion of small molecules in lipid
membranes \cite{neale2011statistical}.  Concurrent to all-atom (AA) models,
coarse-grained (CG) models, which reduce the number of degrees of freedom by
lumping several atoms into larger beads, provide the means to probe essential aspects of these systems and push toward length- and timescales currently unattainable with AA
models \cite{Tschop:1998lj, Kremer:2002ge, peter2009multiscale, Voth:2009,Riniker:2012qf,noid2013perspective}.

Most strategies that aim at parametrizing CG models target static
properties---whether phenomenological or emergent properties in top-down models,
or structural aspects of a reference higher-level simulation in bottom-up
models---while kinetic information is rarely included.  Thus, the resulting CG
simulations display largely uncontrolled dynamics, making any kinetic-based
conclusion difficult to interpret.  More specifically, CG dynamics commonly
exert \emph{faster} behavior compared to AA simulations or experiments, which
can be rationalized by the model's reduced molecular friction
\cite{Harmandaris:2007oy}.  While faster dynamics works to the simulator's
advantage---providing more efficient conformational sampling and
transitions---meaningful dynamical information about the system can only be
extracted if \emph{all} kinetic processes are sped up uniformly or in a
predictable fashion.  While rescaling CG dynamics by a single speed-up factor
has proven useful for individual kinetic observables, e.g., collapse of the
mean-squared displacement of various polymer melts across chain lengths
\cite{Harmandaris:2009oc, salerno2016resolving}, correcting a variety of coupled kinetic processes remains problematic. 

More rigorous approaches to describing accurate dynamics in CG models require
additional parametrization effort.  The Mori-Zwanzig formalism describes the
dynamical evolution of a CG system by means of a generalized Langevin equation,
in which the noise and friction need careful attention to accurately model local
dynamics \cite{zwanzig1961memory, MORI:1965xz}.  Practically, this entails the
parametrization of a friction \emph{tensor} for each CG bead
\cite{Izvekov:2006,Hijon:2010ix,Markutsya:2014,Izvekov:2013,Deichmann:2014}.  Other efforts have focused on emulating friction tensors
via fictitious particles \cite{davtyan2015dynamic}, the estimation of entropy and friction corrections due to coarse-graining \cite{lyubimov2013theoretical}, or choosing the CG mapping to minimize memory \cite{Guttenberg:2013}. 
Clearly, while the static properties of a simulation model only depend on its
force field, its kinetics depend additionally on the applied thermostat---making
the latter's choice and parametrization essential.

Building upon these efforts, we seek a simple and systematic method capable of both analyzing and improving kinetic properties in CG models.  
Specifically, we take two main considerations into account:
\begin{enumerate}
\item Many molecular systems display a variety of kinetic processes at various
  timescales.  Identifying a representative set of kinetic processes and
  selecting relevant observables to probe them requires significant insight into the system.  
Ideally, manual selection of kinetic observables and processes should be avoided.
\item Rigorous approaches to building consistent dynamics in CG models have
  three main disadvantages: ($i$) They are often tedious to parametrize and
  implement (e.g., optimization and numerical integration of friction tensors);
  ($ii$) They scale \emph{down} the accelerated dynamics of CG models to the
  level of the real system---reducing the sampling efficiency closer to AA models.
  From a practical perspective, CG models should ideally exhibit both faster and
  \emph{consistent} dynamics; and lastly ($iii$) The long-timescale effects from assumptions and errors associated with the local kinetic parameters (e.g., friction tensor) can be difficult to control.
\end{enumerate}
Markov state models (MSMs) \cite{Chodera:2006, Noe:2009vn, Bowman:2014}, which
are coarse-grained kinetic models of a simulation trajectory, specifically
address the first point. In particular, by estimating transition probabilities
between predefined microstates, a diagonalization of the resulting transition probability
matrix provides an immediate characterization of the hierarchy of slowest
timescales (i.e., eigenvalues) and associated processes (i.e., eigenvectors).
For systems at equilibrium, the largest eigenvalue (i.e., $\lambda_0 = 1$)
corresponds to an infinite timescale with an associated eigenvector describing the
stationary distribution projected onto the chosen microstates.  Consequently,
MSMs simultaneously characterize the static and kinetic properties of
the system, making them a useful methodology to analyze simulation
models 
\cite{Chodera:2011,Lane:2011,Bowman:2011uq,Buch:2011,Bowman:2012,Plattner:2015,Shukla:2016}.

\begin{figure}[htbp]
  \begin{center}
    \resizebox{\columnwidth}{!}{\includegraphics{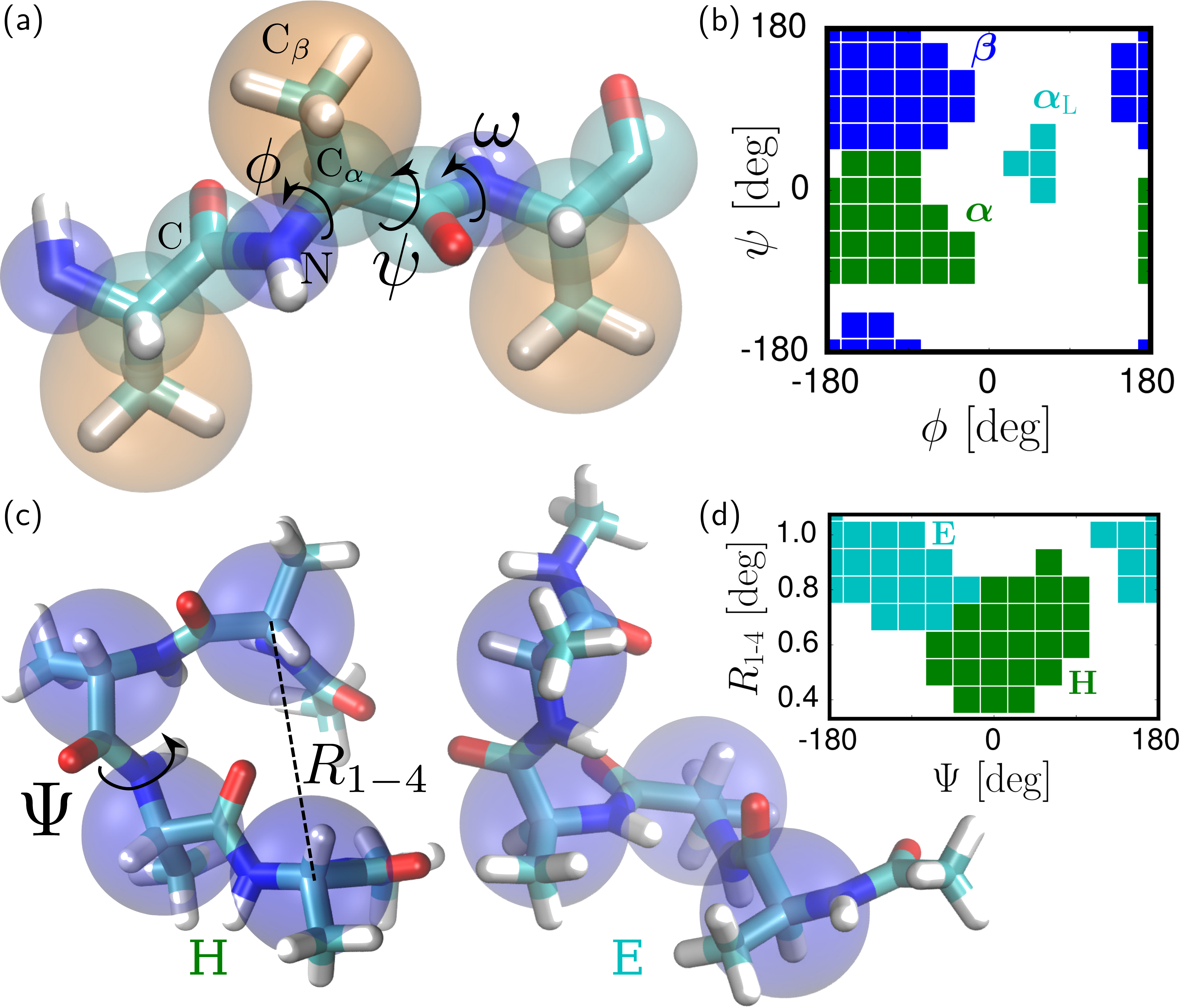} }
    \caption{(a) Cartoon representation of an Ala$_3$ peptide.  Atoms are
      shown using a licorice representation (without terminals), while
      CG beads of the PLUM model are displayed as spheres.  The different
      beads consist of: N for the amide group, C$_\alpha$ for the central
      carbon, C$_\beta$ for the side chain, and C for the carbonyl group.  The
      backbone dihedrals $\phi$, $\psi$, and $\omega$ are depicted as well.
      (b) Illustration of the three relevant metastable states for 
      Ala$_3$ sampled by PLUM,
      projected on a Ramachandran map: helical ($\alpha$), extended
      ($\beta$), and left-handed helix ($\alpha_{\rm L}$) regions. (c) Cartoon
      representation of an Ala$_4$ peptide.  
      Representations of the helical (H) and extended
      (E)  metastable states are shown, as well as the dihedral, $\Psi$, and
      $\of$ residue distance, $\Rof$, order parameters. (d) Illustration of
      the two relevant metastable states for Ala$_4$ sampled by CG-sb, projected onto the two order parameters.  Cartoons rendered with VMD \cite{humphrey1996vmd}.}
    \label{fig:cartoon}       
  \end{center}
\end{figure}

We recently reported the use of MSMs to analyze two CG peptide models---one
top-down and one bottom-up (Figure~\ref{fig:cartoon}) \cite{Rudzinski:2016}.
Peptides offer a useful test for CG models as they display a variety of
kinetic processes at overlapping timescales---even for the smallest of
systems.  We probed to what extent the simulation trajectories were compatible
with more consistent kinetics by building \emph{biased} MSMs, which attempt to
balance the microstate transitions observed in CG simulation trajectories and
reference kinetic information by reweighting the transition probability
matrix.  In this work, we propose refinements to the two peptide models by
altering their force field based on this reweighting.  By focusing on
improving not only the stationary distribution but also ratios of timescales
(i.e., up to an arbitrary speed-up factor) and a qualitative description of the
associated eigenvectors, we use the MSM framework to refine both the static
and kinetic properties of the CG models.  Rather than reference AA
simulations, the present reparametrization relies on biased MSMs for two
reasons: ($i$) this strategy is helpful when no AA simulations are available 
(e.g., but only CG simulations and experimental information) and ($ii$) biased
MSMs can help guide \emph{how} to reparametrize the CG model.  To keep the
parametrization and implementation procedure simple, we limit changes to the
CG models to the beads' masses (i.e., effectively scaling individual isotropic
friction coefficients) and a small number of relevant force-field parameters.
Though these adjustments are not enough to recover a rigorous dynamical
evolution of the systems, we show that they help enhance the quality of the CG
models---not only kinetic, but also static properties.

The present work demonstrates that relying on a combination of static and
kinetic information can help construct more robust CG force fields.  We find
that a simple rescaling of the masses can shift the relative timescales of the
slow kinetic processes, without affecting their nature (i.e., corresponding
eigenvector) or ratio of forward/backward timescales.  As a result, mass rescaling does not necessarily help correct the ordering of the processes.  
Larger alterations of the kinetics require force-field adjustments.  For example, we consider the helix-coil transition sampled by a minimal CG model for a small peptide.  It was previously demonstrated that a structure-based parametrization---aimed at reproducing low-order correlation functions---understabilized the helical state and sampled a large basin of spurious conformational intermediate states \cite{Rudzinski:2014b}.  
We demonstrate here that the inclusion of kinetic
constraints into the parametrization procedure helps identify an alternative
CG model that retains the essential static properties of the original CG model, while improving the ratio of forward/backward timescales by effectively hindering the sampling of these spurious intermediates.

\section{Models and Methodology}

In this work, we investigate the impact of incorporating kinetic information
into the parametrization of CG models.  We consider two small peptide systems
along with two distinct CG approaches: ($i$) a transferable model that retains
resolution of the peptide backbone dihedrals and ($ii$) a highly-specific, structure-based,
1-bead per amino acid model.  For reference, we use
explicitly solvated AA simulations of the corresponding capped peptides, which
employ the OPLS-AA \cite{Jorgensen:1996} and SPC/E \cite{Berendsen:1987} force
fields to model the peptide and solvent interactions, respectively.  These
simulations consisted of a total of 1~$\mu s$ from five independent simulations and were described previously
\cite{Rudzinski:2016,Rudzinski:2014b}.

\subsection{Transferable CG PLUM model applied to Ala$_3$}

\paragraph{Model}
PLUM describes an amino acid using four beads---three for the backbone and one
for the side chain---in an implicit water environment \cite{bereau2009generic}.  
The almost-atomistic
resolution of the backbone allows an explicit modeling of the backbone
dihedrals, $\phi$, $\psi$, and $\omega$ (Figure~\ref{fig:cartoon}).  
While $\omega$---centered around the peptide bond---is largely frozen, $\phi$ and $\psi$ display significant variability and directly map to larger-scale peptide secondary structures.
As such, $\phi$ and $\psi$ are commonly employed in Ramachandran maps \cite{ramachandran1968conformation} to analyze the conformational variability of small peptides.

PLUM was parametrized using a top-down strategy
\cite{noid2013perspective}, which consists of incorporating interactions that
are deemed relevant, and whose parameters are optimized to reproduce emerging
properties of the system.  In PLUM, the parametrization of local interactions
(e.g., sterics) aimed at a qualitative description of Ramachandran maps, while
longer-range interactions---hydrogen bond and hydrophobicitiy---aimed at
reproducing the folding of a three-helix bundle, without explicit bias toward
the native structure \cite{bereau2009generic}.  The model is generic in that
it aims at describing the essential features of a variety of amino-acid
sequences, rather than an accurate reproduction of any specific one.  After
parametrization, it was shown capable to fold several helical peptides
\cite{bereau2009generic, bereau2010interplay, bereau2011structural, 
bereau2014enhanced, bereau2015folding}, stabilize
$\beta$-sheet structures \cite{bereau2009generic, bereau2012coarse,
  osborne2012coarse, osborne2013thermodynamic,
  osborne2014thermodynamics}, and used to probe the conformational
variability of intrinsically disordered proteins \cite{rutter2015testing}.


The kinetics of the PLUM peptide model were probed by means of a
Markov state model analysis of a tri-peptide of alanine residues (Ala$_3$) \cite{Rudzinski:2016}.
Compared with reference AA simulations, the results pointed at significant
issues with the reproduction of the relative timescales of the slowest
processes.  Worse, the CG model switches the order of the two slowest
processes: the transition in and out of the left-handed helical region,
$\alpha_{\rm L}$, and the $\alpha-\beta$ transitions (see
Figure~\ref{fig:cartoon} (b) for a projection of these metastable states on
the Ramachandran map).  The incorporation of reference
mean-first-passage times between the $\alpha$, $\beta$, and $\alpha_{\rm L}$
regions of the Ramachandran map led to a biased Markov state model that
captured the essential features of the CG simulations, while yielding a
realistic hierarchy of kinetic processes and correct timescale separation
\cite{Rudzinski:2016}.  

\paragraph{Simulations}
\label{sec:met:plum}
CG simulations of Ala$_3$ with the PLUM force field \cite{bereau2009generic,
  wang2010systematically, bereau2014more} were run using the {\sc ESPResSo}
simulation package \cite{espresso}.  Details of the force field,
implementation, and simulation parameters can be found in Bereau and Deserno
\cite{bereau2009generic}.  CG units of the model are built from a length,
$\mathcal{L}^{\rm P} = 1$~\AA, an energy $\mathcal{E}^{\rm P} = k_{\rm B}T_{\rm room} \approx
0.6$~kcal/mol as the thermal energy at room temperature, and a mass
$\mathcal{M}^{\rm P} \approx 4.6 \times 10^{-26}$~kg \cite{bereau2009generic}.  The CG
unit of time can be constructed from the combination $\mathcal{T} = \mathcal{L}
\sqrt{\mathcal{M} / \mathcal{E}}$.  Using the values above, we find $\tauplum
\sim 0.1$~ps, which does \emph{not} realistically represent the conformational
changes of the protein \cite{Rudzinski:2016,bereau2009generic}.
A single canonical simulation at temperature $k_{\rm B}T=1.0~\mathcal{E}$ was
performed for $100,000~\tauplum$, recording the system every $0.1~\tauplum$.
Temperature control was ensured by means of a Langevin thermostat with friction
coefficient $\gamma = (1.0~\tauplum)^{-1}$.  

\paragraph{Reparametrization}
For the purpose of the present reparametrization, we first varied the masses,
$m$, of the beads.  While all set to $m = \mathcal{M}$ in the original model
\cite{bereau2009generic}, we considered alternative assignments for the
C$_\alpha$, C$_\beta$, and C beads, while retaining $m_{\rm N} = \mathcal{M}$
(Figure~\ref{fig:cartoon}).  Fixing the mass of one of the beads allows us to
focus on altering the \emph{relative} kinetics, since a uniform shift of all masses merely scales each dynamical process by the same amount.  We
considered alternative masses $m/\mathcal{M} = \{0.1, ~0.5, ~1.0, ~2.0, ~5.0,
~10.0\}$, for a total of $6^3 = 216$ trial models. 

The PLUM force field is made up of a small number of interactions.  
Excluding the bonded interactions, there are steric interactions between beads, the backbone-backbone hydrogen-bond interaction, and side-chain hydrophobicity.
The last two have little effect on the static and kinetic
properties of Ala$_3$, given the small size of the system.
Thus, we focus on the sterics, varying the Weeks-Chandler-Anderson
parameters associated with the bead sizes, $\sigma$, within $10\%$ of their original values (the functional form makes the interactions insensitive
to the strength, $\epsilon$).  For each reparametrization, we ran a canonical simulation and analyzed both the static and kinetic properties using a
Markov state model analysis.

\subsection{Structure-based CG model for Ala$_4$}

\paragraph{Model}
We also considered a system-specific, structure-based implicit solvent CG
model \cite{Rudzinski:2014b} for a tetra-peptide of alanine residues ($\Aq$),
effectively probing a minimal helix-coil transition.  This model, denoted
CG-sb, represents each amino acid with a single CG bead, placed at the
$\alpha$-carbon position.  The potential energy function employs four distinct
interactions between the four CG beads: bond, angle, dihedral and $\of$ (i.e.,
end-to-end).  These potentials were determined using a self-consistent generalized Yvon-Born-Green relation
\cite{Cho:2009ve,Rudzinski:2014} aimed at reproducing a set of force
correlation functions---related to one-dimensional structural
distributions---along each order parameter corresponding to a term in the CG
potential.  The CG-sb model qualitatively reproduces the free-energy surface
along the dihedral angle, $\Psi$, defined between the four $\alpha$-carbons of
the peptide backbone and the end-to-end distance, $\Rof$, between the first
and last $\alpha$-carbons (Figure~\ref{fig:cartoon}c).  
Although CG-sb better stabilizes helices compared to a corresponding force-matching-based model, the minimal representation and simple interactions lead both models to sample a large basin of spurious conformational intermediates, forbidden at the AA level \cite{Rudzinski:2014b}.

The kinetics of the CG-sb model were also recently assessed via Markov state
model analysis \cite{Rudzinski:2016}.  
Compared to reference AA simulations, the CG-sb model correctly assigns
  the transition between helical (H) and extended (E) metastable states as the slowest dynamical process, and with an accurate representation of the
  associated eigenvector (i.e., which microstates are primarily involved in
  the kinetic process).  However, the CG-sb model exhibits transitions from H to E that are too fast compared to the backward process, in line with
  the understabilized helical state \cite{Rudzinski:2014b}.  The model also
  lacks any significant timescale separation between the two slowest dynamical
  processes \cite{Rudzinski:2016}.  
The incorporation of reference mean-first-passage times between the H and E metastable states led to a biased Markov state model that retained the essential features of the original model, while improving the timescale separation and the ratio of forward/backward timescales of the H$-$E transition.  
This improvement was achieved by concentrating the flux of probabilities in a narrow subset of the helical region, as compared to the reference AA trajectory.

\paragraph{Simulations}
CG simulations of $\Aq$ were performed with the Gromacs 4.5.3 simulation suite
in the constant NVT ensemble with $T$ = 298~K, while employing the stochastic
dynamics algorithm with a friction coefficient $\gamma = (2.0~\tausb)^{-1}$ and a time
step of $1e-3~\tausb$, where $\tausb = 1$~ps.
A single simulation was performed for $50,000~\tausb$, recording the system
every $0.05~\tausb$.  The CG unit of time, $\tausb$, can be determined from
estimates of the fundamental units of length, mass, and energy of the simulation
model, but does not provide any meaningful description of the dynamical
processes generated by the model.

\paragraph{Reparametrization}
We considered several reparametrizations of the CG-sb model, which employ the
same representation and set of interaction types as the original model.  The
bond potential remained identical for all models.  
Our reparametrization strategy follows the insight gained from the biased Markov
  state model: vary parts of the CG potentials to better stabilize the helical
  state \cite{Rudzinski:2016}.  As a proxy, we take the difference between the
  CG-sb and force-matching potentials, $\delta U_i$, given the models'
  differing abilities in stabilizing the helix \cite{Rudzinski:2014b}.
  Trial CG potentials were then defined as
  $U^{\operatorname{CG}}_i({\boldsymbol \alpha}) = U^{\operatorname{CG-sb}}_i
  + \alpha_i \delta U_i$, where $i = \{$ang, dih, $\of\}$ adjusts separately
  each angle, dihedral, and $\of$ interaction, respectively.
For each reparametrization, we ran a canonical simulation and analyzed both the static and kinetic properties using a Markov state model analysis.

\subsection{Markov state models}

Given a trajectory generated from molecular dynamics simulations, Markov state
models attempt to approximate the exact dynamical propagator with a finite
transition probability matrix, $\Trm(\tau)$ \cite{Chodera:2006,Noe:2009vn,Bowman:2014}.
This requires a discretization of configuration space, which
groups all possible configurations into a manageable set of microstates.  Once
the microstates are chosen, the number of observed transitions from microstate
$i$ to $j$ at a time separation $\tau$, $C^{\rm obs}_{ij}(\tau)$, is determined.  The
matrix of transition counts, $\Crm^{\rm obs}(\tau)$, embodies the dynamics of
the simulation trajectory.  An estimator for the transition probability matrix
is then constructed such that the simulation data is optimally described, while
ensuring normalization and detailed balance constraints. The latter constraint 
applies to any system at equilibrium and alleviates finite-sampling issues.  The
transition probability matrix is constructed 
 by maximizing the posterior $p(\Trm \mid \Crm) \propto p(\Crm
\mid \Trm) = \prod_{i,j} T_{ij}^{C_{ij}}$
, where we applied Bayes' theorem and
a uniform prior distribution \cite{Prinz:2011b}. 

We recently proposed a method to construct an MSM informed by additional 
external information \cite{Rudzinski:2016}.
By combining the aforementioned optimization procedure with a set of \emph{coarse} reference 
kinetic constraints, we determined a new transition probability matrix 
that best reproduces the constraints while minimally biasing the simulation data.
We refer to such models as biased MSMs.

In the present work, standard MSMs are built for each of the reparameterized CG models, as well as the reference AA models.
Following the projection and discretization of peptide trajectories along the order parameters presented in Figure~\ref{fig:cartoon},  MSMs
were generated via a maximum-likelihood technique \cite{Bowman:2009}.
MSM construction and analysis (e.g., calculation of the eigenspectrum, metastable states, and mean-first-passage times) were performed using the {\sc pyEmma} package \cite{PyEmma,Senne:2012, scherer2015pyemma}.
Lag times of $\tau = 250$~ps and $\tau = 1.25~\mathcal{T}^{\rm S}$ were
employed for the AA and CG models, respectively, for $\Aq$ and $\tau = 40$~ps
and $\tau = 1.5~\mathcal{T}^{\rm P}$ were employed for the AA and CG models,
respectively, for $\At$.\footnote{Rudzinski \emph{et al.} \cite{Rudzinski:2016} mistakenly reported AA lag times in ns instead of ps.}
See the Supporting Information section of \cite{Rudzinski:2016} for more details.

\subsection{Model assessment}
For all reparametrizations considered, 
we systematically compared the results to the biased Markov
state models constructed by Rudzinski \emph{et al.}
\cite{Rudzinski:2016}.  We quantify the similarity between
eigenvectors or stationary distributions of different models using the
Jensen-Shannon divergence
\begin{equation}
  \label{eq:jsd}
  {\rm JSD}(p || q) = \frac 12 \sum_i p_i \left( \ln p_i - \ln m_i \right)
  + \frac 12 \sum_j q_j \left( \ln q_j - \ln m_j \right),
\end{equation}
where $p$ and $q$ are two (discretized) distributions, $i$ and $j$ run over
all bins, and $m = (p+q)/2$.
The JSD provides a symmetrized and smoothed version of the
Kullback-Leibler divergence, $D(p||q) = \sum_i p_i \ln p_i/q_i$
\cite{KULLBACK:1951fk}.

\section{Results}

\subsection{Transferable PLUM model applied to Ala$_3$}

The following describes the refinement of the PLUM model to better describe the statics and kinetics of Ala$_3$.  
As a first step, we monitor to what extent varying the beads' masses can help
improve the kinetic properties of the model.  Given that masses only couple to
the Hamiltonian via the kinetic energy, they do not alter any static equilibrium
property \cite{bereau2011unconstrained}.  Thus, we only monitor the impact on the dominant eigenvalues and eigenvectors of the transition
probability matrix.
Subsequently, we consider larger changes to PLUM by varying force-field components and assess both static and kinetic properties of the resulting models.

\begin{figure}[htbp]
  \begin{center}
    \resizebox{0.9\columnwidth}{!}{\includegraphics{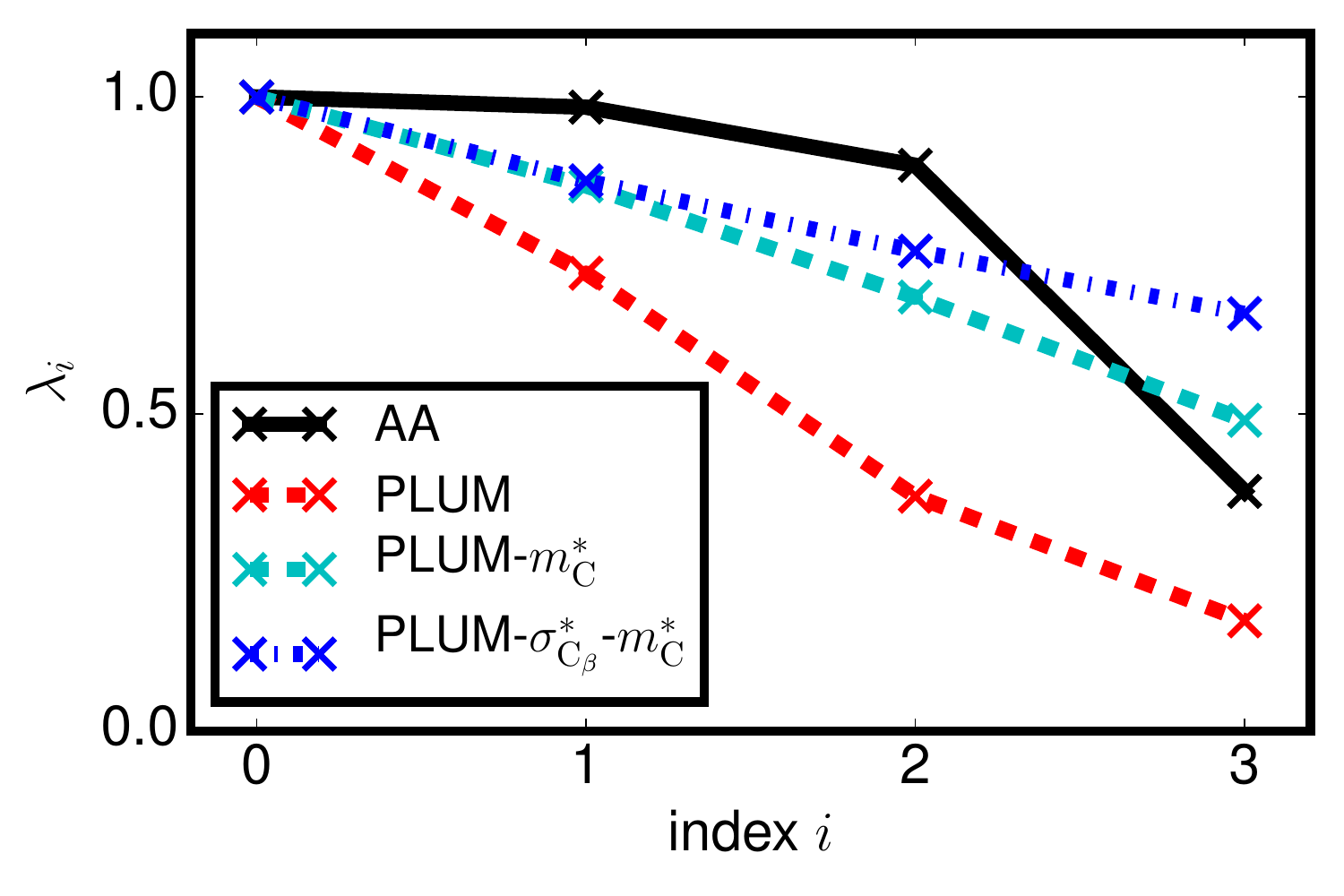} }
    \caption{Ala$_3$: Eigenvalue spectrum of the three slowest processes of
      the reference AA model, the original PLUM model, PLUM with
      rescaled masses (PLUM-$m^*_{\rm C}$), and PLUM with altered force
      field and rescaled masses (PLUM-$\sigma^*_{{\rm C}_\beta}$-$m^*_{\rm
        C}$).  Starting with $\lambda_3$, the corresponding kinetic processes probe transitions \emph{within} metastable states.}
    \label{fig:ala3_evals}       
  \end{center}
\end{figure}

We begin with a brief summary of the kinetic analysis of PLUM applied to $\At$, reported by Rudzinski \emph{et al.} \cite{Rudzinski:2016}.  Reference AA simulations
indicated a strong timescale separation between the second eigenvalue and the
next---singling out the first two kinetic processes as the most relevant.
These correspond to transitions \emph{between} the metastable states depicted
in Figure~\ref{fig:cartoon} (b)---in and out of $\alpha_{\rm L}$ and between $\alpha$ and
$\beta$, respectively---while all other kinetic processes occur \emph{within}
the metastable states.  Although the AA MSM yielded very similar eigenvalues,
$\lambda_2/\lambda_1 \approx 0.9$, the PLUM MSM displayed a first eigenvalue that is too high compared to the second, i.e.,
$\lambda_2 / \lambda_1 \approx 0.4$ (Figure~\ref{fig:ala3_evals}).  Worse, the order of the first two
eigenvectors in PLUM is switched, such that the ratio between $\lambda_1$ and
$\lambda_2$ is qualitatively wrong.

\paragraph{Refinement I: masses}
\label{sec:ala3_m}

\begin{figure}[htbp]
  \begin{center}
    \resizebox{0.95\columnwidth}{!}{\includegraphics{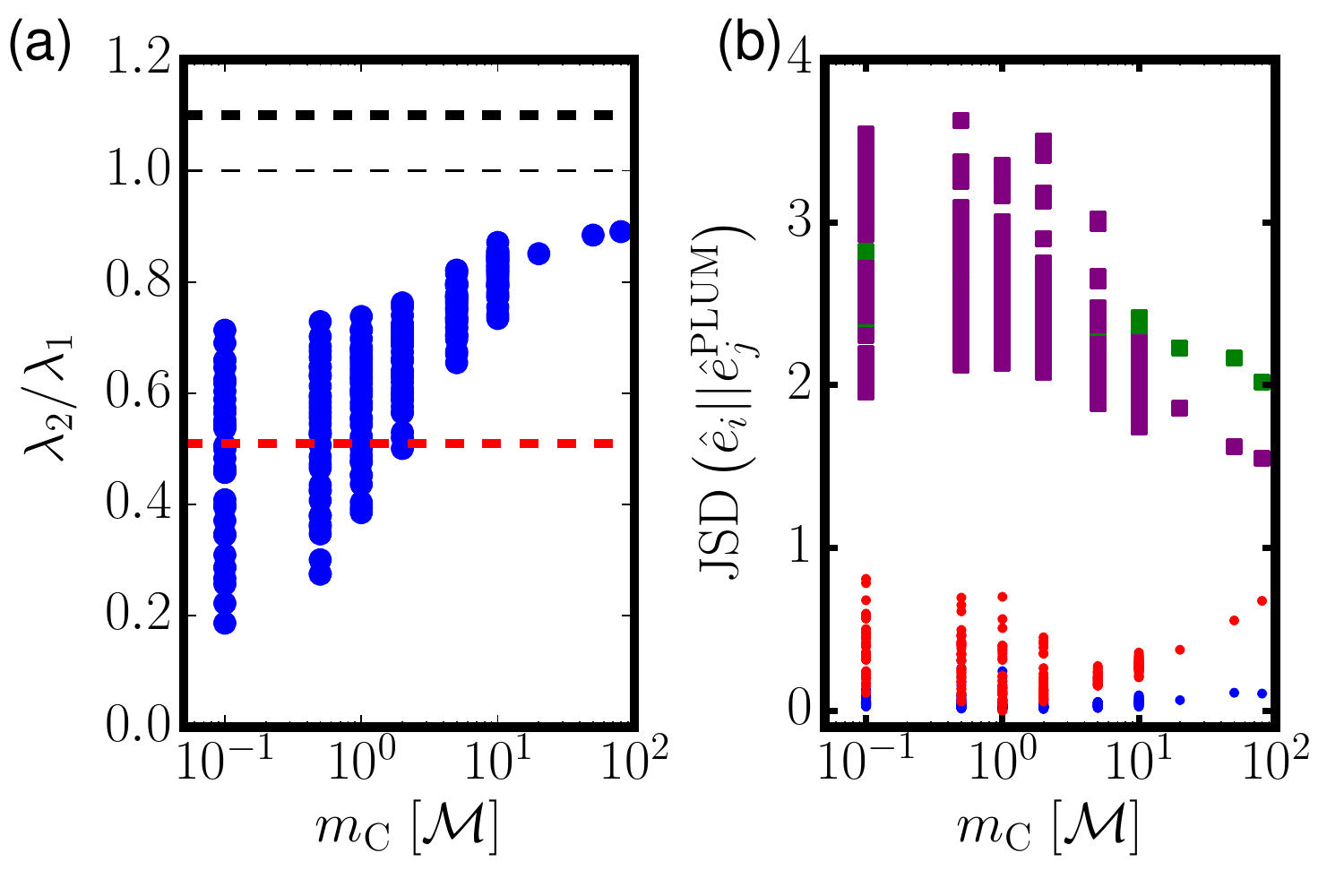} }
    \caption{Ala$_3$: (a) Ratio of the first two eigenvalues
      $\lambda_2/\lambda_1$, indicating the relative timescale of the
      associated kinetic processes.  The bold red and black dashed lines
      indicate values for the original PLUM and reference AA models,
      respectively.  Note that the AA ratio is located above $1.0$ because of
      PLUM's unphysical switching of the kinetic processes.  Blue dots
      correspond to trial CG models with rescaled masses, projected along
      $m_{\rm C}$ while $m_{\rm N} = \mathcal{M}$. (b) Jensen-Shannon
      divergence (equation~\ref{eq:jsd}) between eigenvectors of the trial
      models and the original PLUM model: red and blue dots are between
      eigenvectors of identical index (i.e., $1-1$, $2-2$), while the purple
      and green squares are between eigenvectors of different indices (i.e.,
      $1-2$, $2-1$).}
    \label{fig:ala3_l2l1}       
  \end{center}
\end{figure}

To probe the dependence of the kinetic properties of PLUM's Ala$_3$ on the
masses, we considered a variety of bead-mass combinations as described in the
Methods section.  
Importantly, all models keep $m_{\rm N} = \mathcal{M}$ fixed, such that the effect stems from the \emph{difference} between masses.
For each model, we monitor the accuracy of the \emph{relative} timescales of the first two kinetic processes, $\lambda_2/\lambda_1$, as
shown in Figure~\ref{fig:ala3_l2l1} (a).

Among all beads, we find the strongest variations against $m_{\rm C}$.
Increasing $m_{\rm C}/m_{\rm N}$ is expected to slow processes along $\phi$, since this dihedral directly connects the C beads of consecutive residues.
Kinetically, motion along $\phi$ correlates most strongly with transitions
between the $\alpha_{\rm L}$ and $\beta$ regions (Figure~\ref{fig:cartoon}).
Indeed, we find that increasing $m_{\rm C}/m_{\rm N}$ slows down the second kinetic process in PLUM (i.e., transitions involving $\alpha_{\rm L}$) with respect to the first (i.e., $\alpha-\beta$ transitions), such that $\lambda_2/\lambda_1$ improves (Figure~\ref{fig:ala3_l2l1} (a)).
However, even a ratio $m_{\rm C}/m_{\rm N}$ approaching $10^2$---arguably an unreasonably large difference---does not suffice to reach $\lambda_2 = \lambda_1$, at which point the two eigenvalues switch to correct the order of the kinetic processes.

To further probe the ordering of kinetic processes, Figure~\ref{fig:ala3_l2l1} (b) reports the Jensen-Shannon divergence (JSD;
equation~\ref{eq:jsd}) between eigenvectors of identical index, as well as between
eigenvectors of different indices.   
Overall, the trial simulation models with altered masses yielded similar eigenvectors and none of them corrected the hierarchy of kinetic processes.
The eigenvalue spectrum of an improved mass-rescaling is presented in Figure~\ref{fig:ala3_evals} (PLUM-$m^*_{\rm C}$), for which $m_{\rm C} =
10~\mathcal{M}$ while $m_{{\rm C}_\alpha} = m_{\rm N} = \mathcal{M}$.  The eigenvalue ratio of PLUM-$m^*_{\rm C}$, $\lambda_2/\lambda_1 = 0.8$, is significantly improved compared to the original PLUM model.  
However, mass rescaling does not significantly change the ratio of forward/backward mean-first-passage times between pairs of metastable states (data not shown).  
Altering masses likely scales the dynamics homogeneously across a particular kinetic process, whether forward or backward.  

All in all, these findings suggest that mass rescaling can improve the relative timescale of different kinetic processes, but cannot improve relative mean-first-passage times, and is not necessarily sufficient to correct qualitative discrepancies in the model's
kinetic properties.

\paragraph{Refinement II: force field}

\begin{figure*}[htbp]
  \begin{center}
    \resizebox{0.75\textwidth}{!}{\includegraphics{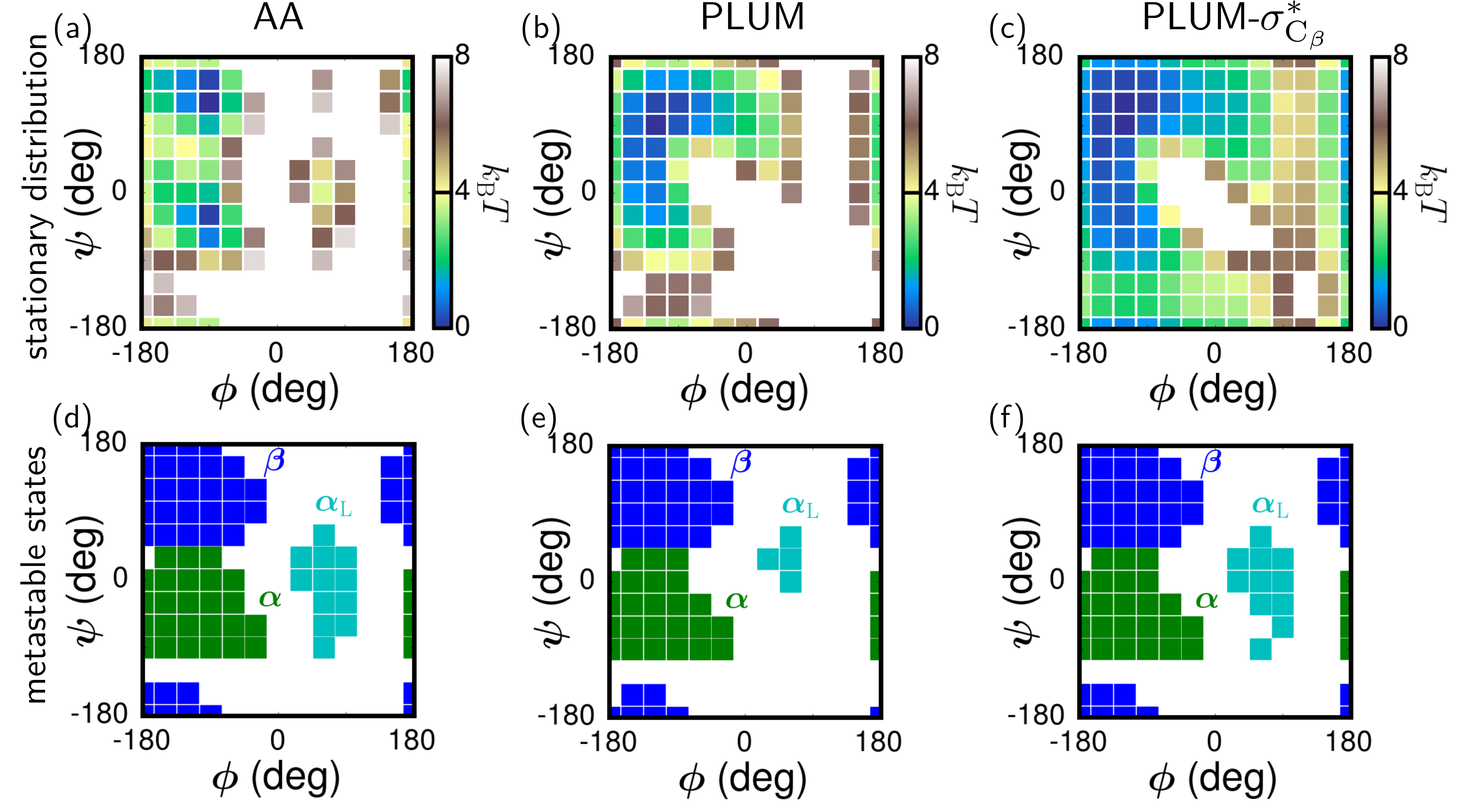} }
    \caption{Ala$_3$: Free-energy surfaces of AA (a), original PLUM (b), and PLUM with smaller side-chain radius, PLUM-$\sigma^*_{{\rm C}_\beta}$ (c).  PLUM-$\sigma^*_{{\rm C}_\beta}$ almost quantitatively reproduces the $\alpha-\beta$ as well as $\alpha_{\rm L}-\beta$ free-energy differences, with respect to the AA model.  Metastable states of the AA model (d) and the corresponding states sampled in the two CG models: (e) PLUM and (f) PLUM-$\sigma^*_{{\rm C}_\beta}$.}
    \label{fig:ala3_sc}       
  \end{center}
\end{figure*}

Though we attempted to vary all bead sizes around their equilibrium values,
the side chain yielded the largest effects on both the static and kinetic
properties.  Specifically, reducing the side-chain bead radius, $\sigma_{{\rm
    C}_\beta}$, by $10\%$ (i.e., from $2.50$~\AA~to $2.25$~\AA) significantly
altered several properties of the force field.  Figure~\ref{fig:ala3_sc} (a--c)
compares the free-energy surfaces of the AA, PLUM, and the altered PLUM force
field with reduced side-chain bead size (PLUM-$\sigma^*_{{\rm C}_\beta}$).  We find a significant stabilization of the sterically-hindered region with $\phi \approx 120~{\rm deg}$ and $\psi \approx -120~{\rm deg}$.  
Though too stable compared to the AA model, the free-energy of this region remains in the range $\approx 2-6~k_{\rm
  B}T$, such that the conformational ensemble at room/body temperature is marginally affected.  Further, this leads to an improved description of the
$\alpha_{\rm L}$ metastable state: the free-energy of $\alpha_{\rm L}$ relative to $\beta$ was previously reported as $4.18$ and $5.75~k_{\rm B}T$ for the AA and original PLUM models \cite{Rudzinski:2016}, while PLUM-$\sigma^*_{{\rm C}_\beta}$ yields $4.10~k_{\rm B}T$, in excellent agreement with AA.  Similarly, the free-energy of $\alpha$ relative to $\beta$ is improved: values of $0.42$, $0.95$, and $0.65~k_{\rm B}T$ are found for the
AA, PLUM, and PLUM-$\sigma^*_{{\rm C}_\beta}$ force fields, respectively.  Figure~\ref{fig:ala3_sc}
(d--f) also displays projections of the AA metastable states, sampled in the CG trajectories. 
We find that the PLUM-$\sigma^*_{{\rm C}_\beta}$ model displays a broader $\alpha_{\rm L}$ region as compared to PLUM, in better agreement with AA.  Kinetic properties are improved slightly, as evidenced by a larger eigenvalue ratio,
$\lambda_2/\lambda_1 = 0.75$, though we obtain eigenvectors similar to PLUM's.

Evidently, the altered force field should not only improve the description of
the Ramachandran map, it ought to retain the features of the original model.
Given that PLUM was parametrized to also reproduce the folding of the \emph{de
  novo} $\alpha$3D three-helix bundle, we verified that PLUM-$\sigma^*_{{\rm C}_\beta}$ could do the same using the original simulation protocol \cite{bereau2009generic}.  We
indeed find spontaneous folding, as monitored by the amount of helicity and
topology of the three helices against the NMR structure
\cite{walsh1999solution}, and a virtually unaltered folding temperature, i.e.,
$k_{\rm B}T_f \approx 1.2~\mathcal{E}$ (data not shown).

Finally, applying the mass rescaling proposed above,
we obtain a model (PLUM-$\sigma^*_{{\rm C}_\beta}$-$m^*_{\rm C}$) that further increases the eigenvalue ratio,
$\lambda_2/\lambda_1 = 0.87$, only $20\%$ off from the AA ratio, while the
original PLUM model was more than $50\%$ too low (Figure~\ref{fig:ala3_evals}).  In
terms of mean-first-passage times, PLUM-$\sigma^*_{{\rm C}_\beta}$-$m^*_{\rm C}$ slightly improves the forward/backward timescale ratio for transitions between the metastable states (data not shown).  

\subsection{Structure-based model for Ala$_4$}

\begin{figure*}[htbp]
  \begin{center}
    \resizebox{0.65\textwidth}{!}{\includegraphics{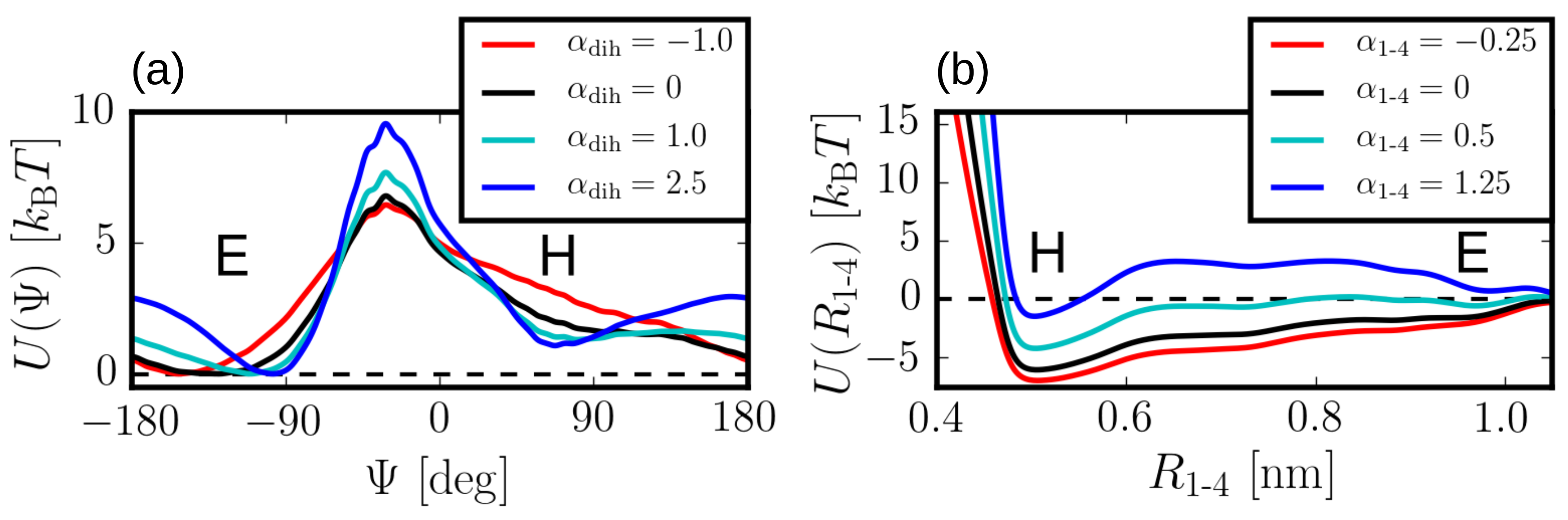} }
    \caption{$\Aq$: Reparametrizations of the dihedral (a) and $\of$ (b) potentials for the CG-sb model of $\Aq$ as a function of the parameter $\boldsymbol \alpha = \{ \alpha_{\rm ang}, \alpha_{\rm dih}, \alpha_{\rm \of} \}$.  The black curves correspond to the CG-sb model ($\boldsymbol \alpha = \{0,0,0\}$), while the cyan curves correspond to the reparametrized CG* model ($\boldsymbol \alpha = \{1.0,1.0,0.5\}$). The H and E labels refer to the helical and extended metastable states, respectively (see Figure~\ref{fig:cartoon} (d)).}
    \label{fig:ala4_params}       
  \end{center}
\end{figure*}

We also assessed refinements of the CG-sb model to better reproduce features of the helix-coil transition for $\Aq$.
We consider variations in the angle, dihedral, and $\of$
potentials of the CG-sb model according to individual scaling parameters, $\alpha_i$, for each interaction $i$.  
Figure~\ref{fig:ala4_params} illustrates characteristics of the resulting interaction potentials for the dihedral and $\of$ degrees of freedom.  
Panel (a) demonstrates that the minima in the dihedral interaction corresponding to the helix 
(H, $\Psi \approx 60$ deg) and extended (E, $\Psi \approx -120$ deg) regions become increasingly
narrow as $\alpha_{\rm dih}$ shifts from $-1$ (red curve) to $2.5$ (blue curve).   
The E minimum also shifts significantly to the right---a limitation of our parameter search.  
In addition to the narrowing, the H region is stabilized relative to E.  
Similarly, panel (b) demonstrates that the
H minimum ($\Rof \approx 0.5$ nm) in the $\of$ interaction is both narrowed and destabilized as
$\alpha_{\of}$ shifts from $-0.25$ (red curve) to $1.25$ (blue curve).
The black curve in each panel of Figure~\ref{fig:ala4_params} denotes the interaction in the CG-sb model. 
Finally, adjustments of the angle parameter ($0 < \alpha_{\rm ang} < 1.5$)
tilted the potential to stabilize the H region relative to E.  

When considering trial reparametrizations, we monitor both the reproduction of the kinetic properties against the biased MSM \cite{Rudzinski:2016} and structural properties compared to the original parametrization \cite{Rudzinski:2014b}.
In particular, improvements in the kinetics were assessed from the Jensen-Shannon divergence (JSD; equation~\ref{eq:jsd}) of the first dynamical eigenvector, $\hat{\boldsymbol e}_1$, with respect to the biased MSM, while divergence from the structural properties of the CG-sb model was quantified with the JSD of the stationary distribution, $\boldsymbol \pi$.

\begin{figure*}[htbp]
  \begin{center}
    \resizebox{0.65\textwidth}{!}{\includegraphics{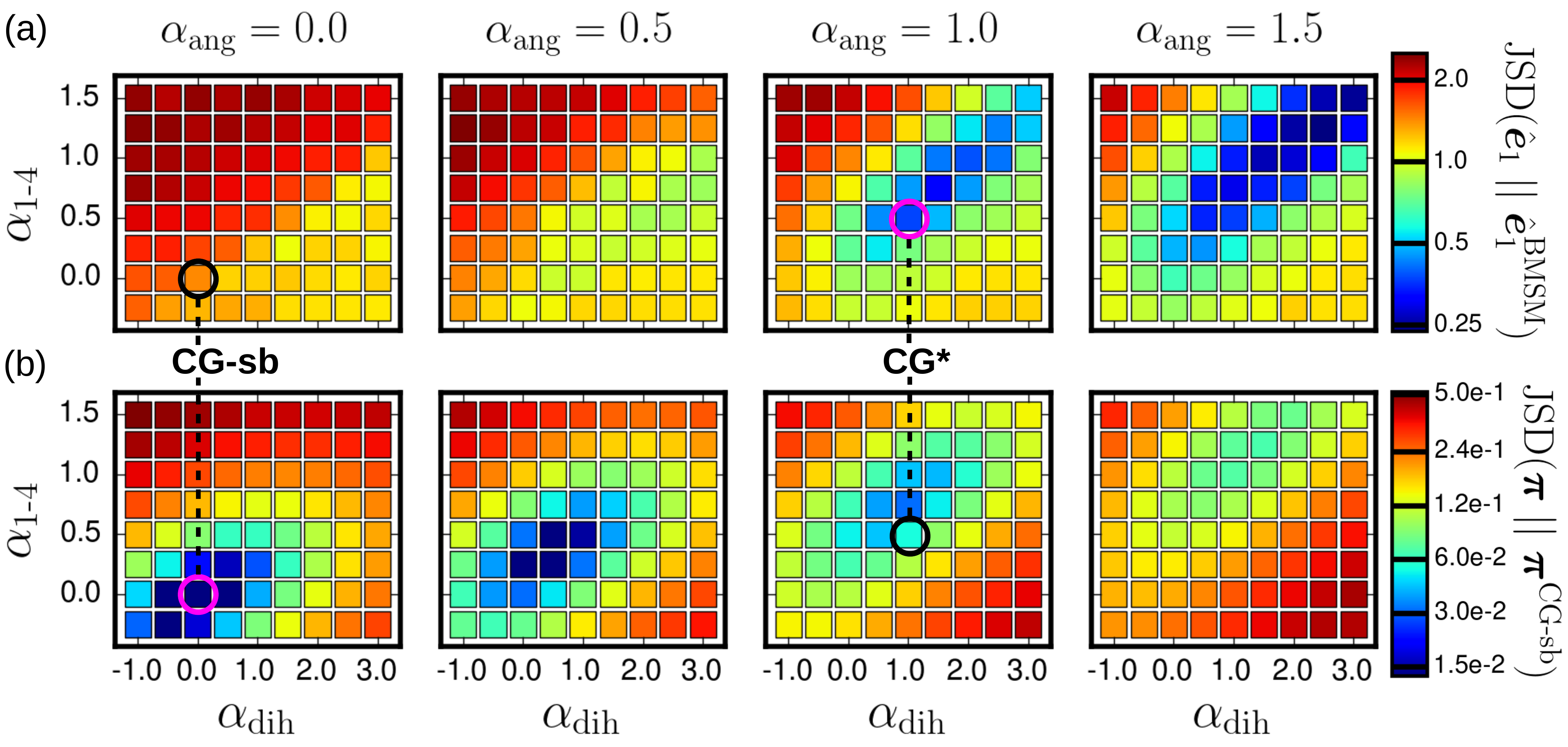} }
    \caption{$\Aq$: Assessment of the reparametrized models, each characterized by the
      scaling parameter $\boldsymbol \alpha = \{ \alpha_{\rm ang}, \alpha_{\rm dih},
      \alpha_{\rm \of} \}$ (See Figure~\ref{fig:ala4_params}). (a) Kinetic
      discrepancies are characterized with respect to the biased MSM (BMSM)---representing the minimal adjustments to the CG-sb model that will
      yield consistent kinetics---by the Jensen-Shannon divergence ($\rm JSD$) of the
      first dynamical eigenvector: $\JSDeone$.  (b) Structural discrepancies are
      characterized with respect to the original CG-sb model---parametrized with
      a structure-based scheme---by the JSD of the stationary distribution,
      $\boldsymbol \pi$: $\JSDpi$). Note that the global minimum of $\JSDpi$, $\boldsymbol \alpha = \{ 0,0,0 \}$, falls below the plotted scale ($\approx 10^{-4}$) and is not exactly equal to zero due to small uncertainties in the construction of the MSM. }
    \label{fig:ala4_param_search}       
  \end{center}
\end{figure*}
 
Columns 1 through 4 of Figure~\ref{fig:ala4_param_search} characterize these two metrics for increasing values of the angle reparametrization parameter, $\alpha_{\rm ang}$, as a function of 
$\alpha_{\rm dih}$ and $\alpha_{\of}$ (see Figure~\ref{fig:ala4_params}).
Figure~\ref{fig:ala4_param_search} (a) demonstrates that the CG-sb model
($\boldsymbol \alpha = \{0,0,0\}$) lies away from any local minimum of
$\JSDeone$.  Increased values of $\boldsymbol \alpha$ tend to improve the kinetics
relative to the CG-sb model.  
Figure~\ref{fig:ala4_param_search} (b) demonstrates that, while changing $\boldsymbol
\alpha$ necessarily increases $\JSDpi$, there exists a slow direction of divergence (i.e., approximately along $\delta\boldsymbol \alpha =
\{c,c,c/2\}$, for $c > 0$). 
Using these metrics, we identify a set of parameters that yields a compromise between kinetics and statics.
We select $\boldsymbol \alpha = \{1.0,1.0,0.5\}$ (denoted CG*) for further analysis below.

\begin{figure*}[htbp]
  \begin{center}
    \resizebox{0.6\textwidth}{!}{\includegraphics{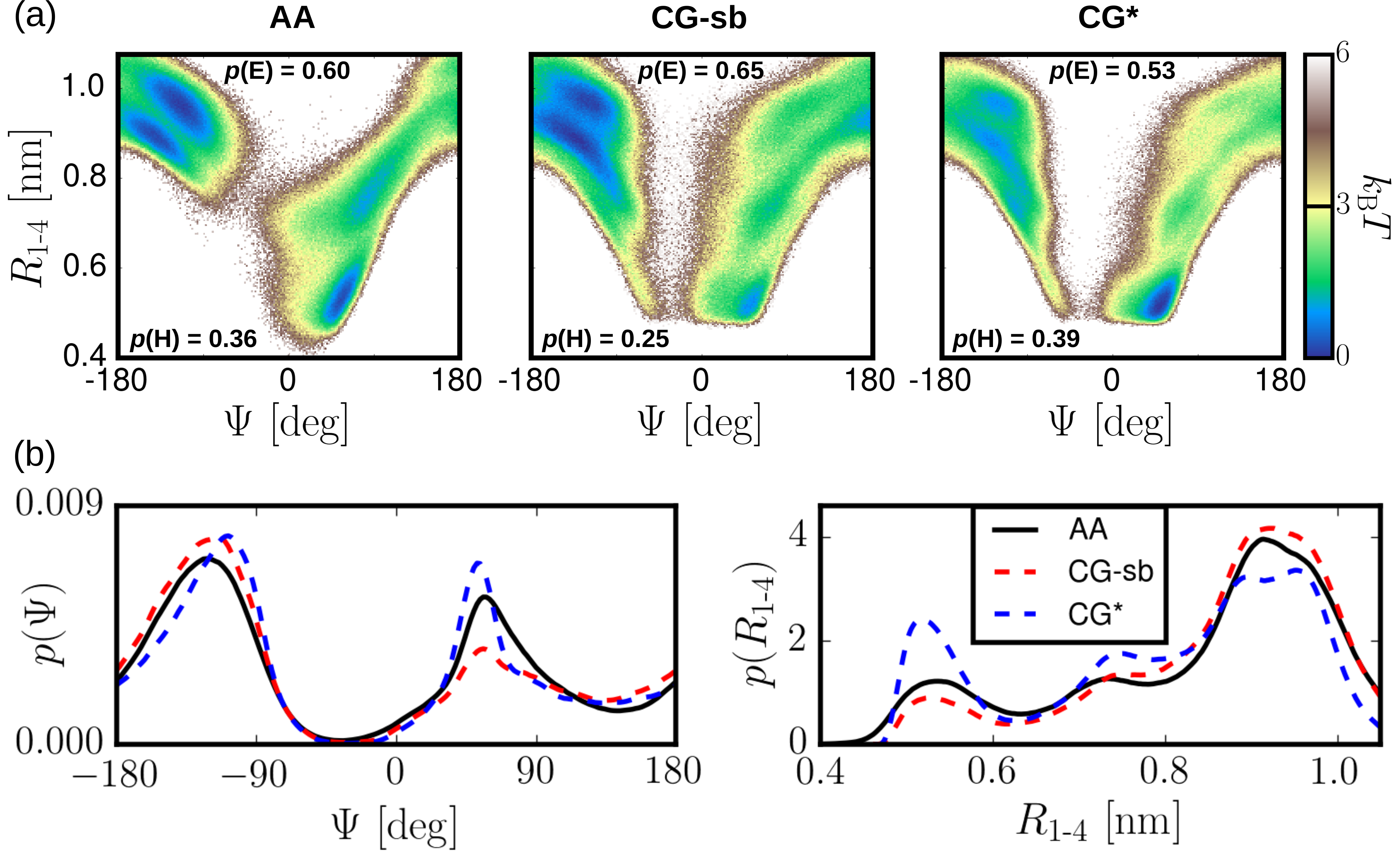} }
    \caption{$\Aq$: Structural characterization of the CG* model with respect to the
      original CG-sb model as well as the reference AA model.  (a) Free-energy surfaces along
      $\Psi$ and $\Rof$. Probability of sampling the H and E metastable states
      (see Figure~\ref{fig:cartoon}) is quantified in each panel. (b) One-dimensional
      distribution functions along $\Psi$ and $\Rof$.}
    \label{fig:ala4_struct}       
  \end{center}
\end{figure*}

In the following, we analyze the structural and kinetic properties of the CG* model with respect to the CG-sb and reference AA models.
Figure~\ref{fig:ala4_struct} (a) presents the free-energy surface of each model along $\Psi$ and $\Rof$.
CG* demonstrates a stabilized H region, $p({\rm H}) = 0.39$, compared with the CG-sb model, $p({\rm H}) = 0.25$, in better agreement with the AA model, $p({\rm H}) = 0.36$.
Both CG models sample the E region with comparable accuracy.
The minimal CG representation (i.e., mapping and interaction set) prevents a quantitative reproduction of the AA cross-correlations, such that both CG-sb and a corresponding force-matching model demonstrate difficulties in stabilizing helices \cite{Rudzinski:2014b}.  This deficiency was linked to the stabilization of spurious intermediates in both models.  
Critically, the CG* model samples fewer of these intermediates ($\Psi \approx 70$~deg, $\Rof \approx 0.95$~nm), compared with CG-sb. We find that models with lower $\JSDeone$ more significantly prohibit this region.
Figure~\ref{fig:ala4_struct} (b) presents the one-dimensional distribution 
functions along $\Psi$ and $\Rof$.
The CG-sb model provides an improved description of the one-dimensional distributions, in comparison to the force-matching model, by sharpening the minima of the CG potentials \cite{Rudzinski:2014b}.
Overall, the CG* model emphasizes the helical region, but retains a qualitative description of the distributions, as compared to CG-sb.

\begin{figure*}[htbp]
  \begin{center}
    \resizebox{0.6\textwidth}{!}{\includegraphics{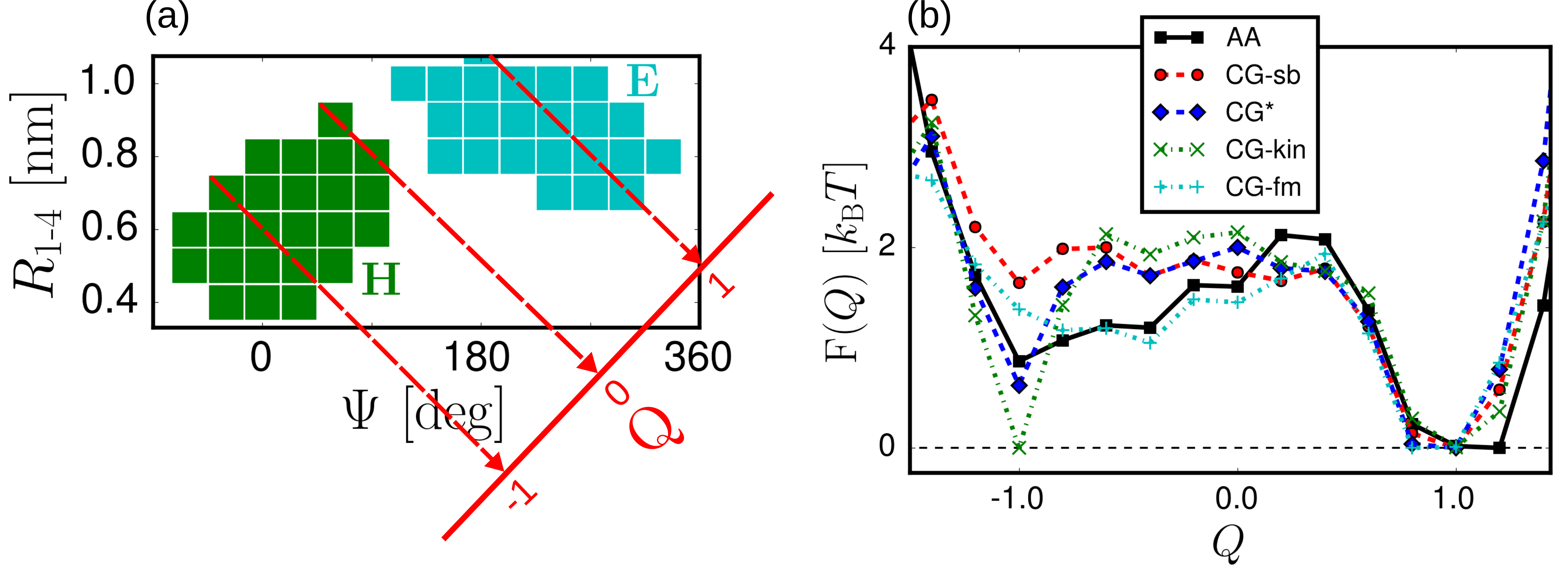} }
    \caption{$\Aq$: Projection of the helix-coil transition in $\Aq$ onto a single order parameter, $Q$.  (a)  Schematic of the projection of the metastable states onto $Q$.  (b) Free-energy profile along $Q$ for the AA model (black curve) and various CG models.}
    \label{fig:ala4_projFES}       
  \end{center}
\end{figure*}

To better monitor the impact of the intermediate region on the nature of the helix-coil transition, we project the free-energy surface along a single parameter, $Q$, defined orthogonal to the dividing surface between metastable states (Figure~\ref{fig:ala4_projFES}).
The black curve demonstrates that the H to E transition in the AA model occurs via a wide and shallow landscape in the H region, with a narrow but steep dividing surface ($Q \approx 0.3$).
Interestingly, the force-matching model constructed by Rudzinski and Noid \cite{Rudzinski:2014b} (CG-fm, cyan curve) reproduces the free-energy along $Q$ rather well, although the lack of a clear minimum in the H region causes serious discrepancies in the one-dimensional distributions and free-energy surface along $\Psi$ and $\Rof$.
In contrast, the CG-sb model (red curve) stabilizes helices via a shallow but narrow minimum in a sub-region of H ($Q \approx -1.0$).
Clearly, this shallow minimum along with the lack of a significant barrier between the states causes exceedingly fast H to E transitions, compared with the reverse process \cite{Rudzinski:2016}.
Relative to the CG-sb model, the CG* model (blue curve) significantly stabilizes the narrow helical minimum, while leaving the remainder of the free-energy profile largely unchanged.
Consequently, the effective barrier between H and E regions is much higher.
Models that emphasize kinetic over structural accuracy exaggerate this stabilization, e.g., CG-kin, $\boldsymbol \alpha = \{ 1.5, 1.0, 1.0 \}$, green curve of Figure~\ref{fig:ala4_projFES}. 
In agreement with the biased MSM results, improved kinetics of the CG model require a sharp and deep minimum in the helical region.

\begin{figure*}[htbp]
  \begin{center}
    \resizebox{0.6\textwidth}{!}{\includegraphics{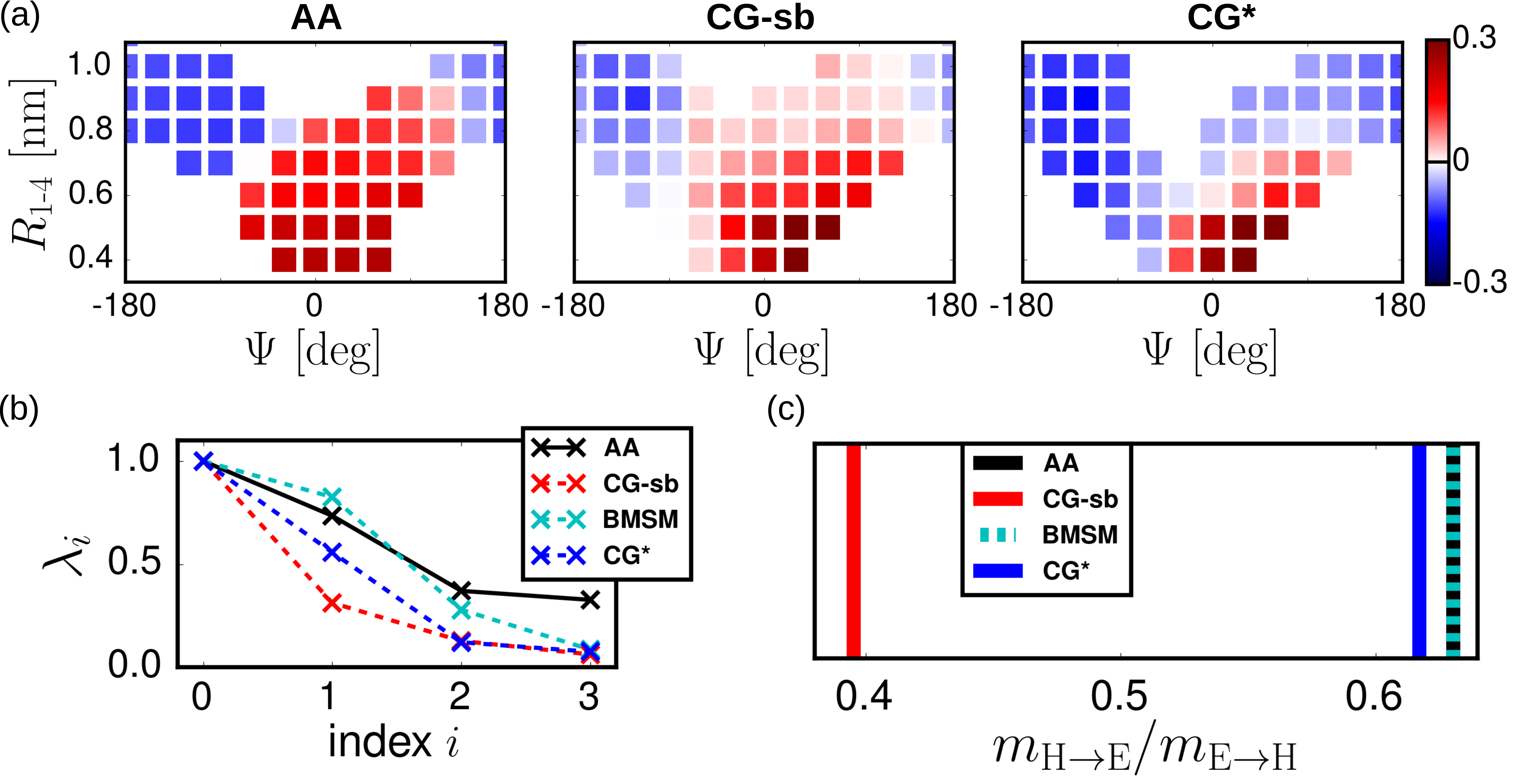} }
    \caption{$\Aq$: Kinetic characterization of the CG* model, compared with the CG-sb model, with respect to the biased MSM (BMSM) and reference AA models. (a) Intensity plots of the first dynamical eigenvectors, characterizing the probability flux of each microstate at the corresponding timescale. (b) Eigenvalue spectrum of the three slowest processes. (c) Ratio of mean-first-passage times between the H and E metastable states.}
    \label{fig:ala4_kin}       
  \end{center}
\end{figure*}

Finally, Figure~\ref{fig:ala4_kin} characterizes the kinetic properties of the CG-sb and CG* models with respect to the biased MSM and reference AA models.
Figure~\ref{fig:ala4_kin} (a) presents an intensity plot of the eigenvector corresponding to $\lambda_1$, which characterizes the probability flux at the corresponding timescale.
The CG-sb model describes the process quite well, albeit with a slight narrowing of the flux in a small region of the H state \cite{Rudzinski:2016}.  
The CG* model demonstrates a similar, but slightly exaggerated, behavior in its eigenvector---a predicted characteristic for consistent kinetics \cite{Rudzinski:2016}.
Figure~\ref{fig:ala4_kin} (b) shows an improved timescale separation between
the first and second kinetic processes in the CG* model compared with the
CG-sb model.  
To further assess the description of the slowest process,
Figure~\ref{fig:ala4_kin} (c) presents the ratio of mean-first-passage times between
the two metastable states.  
The biased MSM reproduces the AA ratio by construction, while the CG-sb yields an H to E transition which is too fast by about 35$\%$,
compared with the reverse process.  
The CG* model displays significant improvement, nearly quantitatively reproducing this ratio.

\section{Conclusions}

The vast majority of coarse-grained (CG) models display grossly inaccurate kinetic 
properties---not only is the dynamics overall faster, the slow kinetic processes 
display inconsistent speedups.  Systems as simple as small polypeptides serve as excellent
examples: from incompatible forward- and backward-rates (e.g., Ala$_4$ above) to
a swapped hierarchy of the slowest kinetic processes (e.g., Ala$_3$ above). Rather than
a rigorous evolution of CG dynamics, we explore to what extent adjustments of the force
field, coupled to a simple Langevin thermostat, can alleviate the issues brought
forward by a Markov state model analysis of the statics and kinetics.  In particular,
the use of \emph{biased} Markov state models, a simulation trajectory augmented with
coarse reference information, provides us with two key advantages: ($i$) the use of
experimental kinetic information can bypass the requirement to rely on expensive and
potentially inaccurate AA simulations; and ($ii$) biased Markov state models
can hint at a reparametrization strategy by highlighting the CG potentials'
deficiencies.

We find that tuning the difference between CG bead masses can improve the relative timescales of the slow kinetic processes, but does not improve the description of the associated eigenvectors and fails to correct the \emph{order} 
of kinetic processes.  
Adjusting the side-chain's bead size of the transferable CG PLUM model can significantly improve the description of the low-populated 
left-handed helix region in Ala$_3$, as seen from the relative stability of the different metastable states (i.e., $\alpha$, $\beta$, and $\alpha_{\rm L}$).
Importantly, we find that the force-field parametrization is still capable of
folding the three-helix bundle $\alpha$3D at a folding temperature in agreement with the original parametrization \cite{bereau2009generic}.  The system was too small to probe and possibly refine the interactions modeling hydrogen bonds and hydrophobicity.

Force-field refinement of the structure-based Ala$_4$ CG model proved most insightful.
The original parametrization, aimed to reproduce one-dimensional correlation functions \cite{Rudzinski:2014b}, lacked stabilization of the helical state, partially due to spurious intermediates forbidden at the AA level. 
Here, we find that alternative force-field parameters targeting both static and kinetic properties can retain the overall quality of the one-dimensional distribution functions, while improving the consistency of transitions between metastable states.  
More specifically, kinetic constraints can restrict the stability of the connecting intermediates, yielding an improved description of transitions.

Although CG potentials that generate a particular set of one-dimensional distributions are unique in theory \cite{Henderson:1974, Chayes:1984a, Rudzinski:2011}, the force field parameters are often highly degenerate in practice \cite{Andersen:1976,MullerPlathe:2002,Rudzinski:2012vn} (i.e., many distinct potentials may give rise to \emph{nearly} identical distributions).
Thus, similar to previous approaches that couple structure-based schemes with additional (e.g., thermodynamic \cite{Ganguly:2012,Dunn:2015}) constraints, consideration of kinetic information can assist in building more robust CG models.
Additionally, matching low order distribution functions alone often explicitly deteriorates the description of cross-correlations \cite{Rudzinski:2014}, which may be generally important for accurately modeling the hierarchical structures stabilized by many biological molecules (e.g., proteins).
Consequently, we expect force-field parametrization strategies that combine static and  kinetic properties to be of use to simulators, even when probing static properties alone.

\section{Acknowledgements}
  We thank Anton Melnyk and Svenja W\"orner for critical reading of the manuscript.  We are also very grateful to Will Noid for the use of the $\Aq$ CG models and simulation trajectories.  Funding from the SFB-TRR146 grant (JFR) and an Emmy Noether fellowship (TB) of the German
  Research Foundation (DFG), as well as a Humboldt Fellowship (JFR) from the
  Alexander von Humboldt Foundation are gratefully acknowledged.  We dedicate
  this work to Kurt Kremer, on the occasion of his 60th birthday, whose
  guidance, advice, and encouragements foster an enriching scientific
  experience within the Theory Group of the Max Planck Institute for Polymer
  Research.

\bibliographystyle{spphys} 

\end{document}